\documentclass[aps,prd,showpacs,preprintnumbers,superscriptaddress,nofootinbib,twocolumn]{revtex4}%
\usepackage[dvips]{graphicx}
\usepackage{bm,latexsym,amsmath,amssymb,amsfonts}
\usepackage{color}
\input{colordvi.tex}

\newcommand*{\D}{{\rm d}}

\newcommand*{\cN}{{\cal N}}
\begin{document}

\title{Large scale evolution
of the curvature perturbation in Ho\v{r}ava-Lifshitz cosmology}

\author{Tsutomu~Kobayashi}
\email[Email: ]{tsutomu"at"gravity.phys.waseda.ac.jp}
\affiliation{Department of Physics, Waseda University, Okubo 3-4-1, Shinjuku, Tokyo 169-8555, Japan}
\author{Yuko~Urakawa}
\email[Email: ]{yuko"at"gravity.phys.waseda.ac.jp}
\affiliation{Department of Physics, Waseda University, Okubo 3-4-1, Shinjuku, Tokyo 169-8555, Japan}
\author{Masahide~Yamaguchi}
\email[Email: ]{gucci"at"phys.aoyama.ac.jp}
\affiliation{Department of Physics and Mathematics, Aoyama Gakuin University,
Sagamihara 229-8558, Japan}

\begin{abstract}
In the non-relativistic theory of gravity recently proposed by
Ho\v{r}ava, the Hamiltonian constraint is not satisfied locally at each
point in space.
The absence of the local Hamiltonian constraint allows the system to have
an extra dark-matter-like component as an integration constant.
We discuss consequences of this fact in the context of
cosmological perturbations, paying a particular attention to
the large scale evolution of the curvature perturbation.
The curvature perturbation is defined in a gauge invariant manner
with this ``dark matter'' taken into account.
We then clarify the conditions under which the curvature perturbation
is conserved on large scales.
This is done by using the evolution equations.
\end{abstract}

\preprint{WU-AP/304/09}
\maketitle

\section{Introduction}

A power-counting renormalizable theory of gravity, proposed recently by
Ho\v{r}ava~\cite{Horava:2008ih, Horava}, has attracted much attention.  The essential
aspect of the theory is broken Lorentz invariance in the ultraviolet
(UV), where it exhibits a Lifshitz-like anisotropic scaling, $t\to
\ell^z t$, $\Vec{x}\to\ell\Vec{x}$, with the dynamical critical exponent
$z=3$.  This will bring an interesting change in the physics of the
early universe since the UV effect may play an important role there.
The study of the cosmology based on Ho\v{r}ava gravity, which is called
Ho\v{r}ava-Lifshitz (HL) cosmology, has been initiated by
Refs.~\cite{Calcagni, Kiritsis}, and since then various aspects of HL
cosmology have been explored, including the generation of chiral
gravitational waves~\cite{TakahashiSoda}, a new mechanism to generate a
scale invariant primordial spectrum without
inflation~\cite{Mukohyama-scale-inv}, the bouncing
scenario~\cite{Bounce}, and others~\cite{Aspects}.  Aside from
cosmology, other interesting works can also be found in~\cite{OtherHL}.

Several versions of Ho\v{r}ava gravity have been known, which are
classified according to whether or not the detailed balance and the
projectability conditions are imposed.  Among them the theory with
projectability and without detailed balance is argued to evade the
problems~\cite{Mukohyama1, Mukohyama2} pointed out in the
literature~\cite{Cai:2009dx, Charmousis:2009tc, Li:2009bg,
Blas:2009yd,Kobakhidze:2009zr} (see also~\cite{Bogdanos:2009uj}).
The most distinguished feature of projectable Ho\v{r}ava gravity is that the
Hamiltonian constraint is not a local equation satisfied at each point
in space, but rather a global equation integrated over the whole space.
Since the global Hamiltonian constraint is less restrictive than the
local one, it allows for a wider class of solutions which contain an
additional dust-like component as an integration constant, as was
clearly remarked in~\cite{Mukohyama1}.


In this paper, we discuss consequences of the absence of the local
Hamiltonian constraint within the context of cosmological perturbations.
Cosmological perturbations in Ho\v{r}ava gravity have already been
investigated in~\cite{Piao:2009ax, Gao:2009bx, Chen:2009jr,
GaoBrandenberger, Cai:2009hc, Koh, YKN, W-M, Lu:2009he}, many of which
are interested in the modified dispersion relation in the UV regime.
Our focus here is on the large scale evolution of the curvature
perturbation in the Sotiriou-Visser-Weinfurtner generalization of
Ho\v{r}ava gravity~\cite{Sotiriou}.
Wang and Maartens were the first to study cosmological perturbations
in this version of the theory~\cite{W-M}.
In general relativity, the curvature perturbation on
uniform density hypersurfaces, commonly denoted as $\zeta$, is conserved
on large scales, provided that the non-adiabatic pressure perturbation
is negligible. This fact can be proven by utilizing the energy
conservation law only~\cite{LythWands}, and hence the conservation of $\zeta$
holds true in a wide range of gravity theories such as brane-world gravity~\cite{brane-world}.
In Ho\v{r}ava gravity, however,
an extra degree of freedom mimicking dark matter,
or, what is dubbed as
``dark matter as an integration constant'' in~\cite{Mukohyama1, Mukohyama2},
appears as a natural outcome of the lack of the local Hamiltonian
constraint.
This forces one to consider effectively a multi-fluid system
even if the system is composed of a single (real) fluid, which implies that
the effect of the entropy perturbation may not be negligible.
Furthermore, individual components are in general not conserved separately,
while the total energy including ``dark matter'' thus introduced
is shown to be conserved locally (by invoking the evolution equations).
Therefore, we start with defining the gauge invariant curvature
perturbation, emphasizing the projectability condition and
taking into account the presence of
``dark matter as an integration constant.''
Then, we discuss the conditions under which the curvature perturbation
is conserved on large scales by using the evolution equations.

This paper is organized as follows. In the next section, the basic
equations in Ho\v{r}ava's non-relativistic theory of gravity is
provided. We review the background cosmology in Ho\v{r}ava gravity in
Sec.~III, emphasizing the consequence of the absence of the local
Hamiltonian constraint. Then, in Sec.~IV, we study the large scale
evolution of the curvature perturbation in the absence of the local
Hamiltonian constraint and clarify the conditions under which it is
conserved on large scales.
We draw our conclusions in Sec.~V.

\section{Ho\v{r}ava gravity}

We consider the projectable version of Ho\v{r}ava gravity without
detailed balance.  The dynamical variables are ${\cal N}$, ${\cal N}_i$,
and $\gamma_{ij}$, in terms of which the metric can be written as
\begin{eqnarray}
\D s^2 =-{\cal N}^2\D t^2 +\gamma_{ij}\left(\D x^i+{\cal N}^i \D t\right)
\left(\D x^j+{\cal N}^j \D t\right),
\end{eqnarray}
with ${\cal N}^i = \gamma^{ij}{\cal N}_j$.  The projectability condition
states that the lapse function depends only on the time coordinate,
${\cal N}={\cal N}(t)$, while ${\cal N}_i$ and $\gamma_{ij}$ may depend
on $t$ and $\Vec{x}$.  The theory is invariant under the
foliation-preserving diffeomorphism: $t\to \tilde t(t)$, $x^i \to
\tilde{x}^i(t, \Vec{x})$.  Under the infinitesimal transformation,
\begin{eqnarray}
t\to t + \chi^0(t), \quad
x^i \to x^i + \chi^i(t, \Vec{x}),
\label{gtrans}
\end{eqnarray}
these variables transform as
\begin{eqnarray}
\gamma_{ij}&\to&\gamma_{ij}-\dot{\gamma}_{ij} \chi^0 - 
\gamma_{ik}\nabla_j\chi^k -\gamma_{jk}\nabla_i\chi^k,
\nonumber\\
{\cal N}&\to&{\cal N} - {\cal N}\dot\chi^0-\dot{\cal N} \chi^0,
\label{trans}\\
{\cal N}_i&\to&{\cal N}_i -
\nabla_i\chi^j{\cal N}_j-\chi^j\nabla_j{\cal N}_i-\dot{\chi}^j\gamma_{ij}
-\dot{\chi}^0{\cal N}_i-\chi^0\dot{{\cal N}}_i,
\nonumber
\end{eqnarray}
where a dot is the derivative with respect to the time coordinate $t$
and $\nabla_i$ the covariant derivative associated with the spatial
metric $\gamma_{ij}$. One can see that ${\cal N}$ remains
$\Vec{x}$-independent after the transformation, and thus it is natural
to impose the projectability condition.

The dynamical variables are subject to the action~\cite{Horava, Sotiriou}
\begin{eqnarray}
S & =& \frac{1}{16\pi G}
\int \D t\D^3x {\cal N}\sqrt{\gamma}
\left(K_{ij}K^{ij}-\lambda K^2
+ R +{\cal L}_{V 2} \right)
\nonumber\\
&&\quad
+\int\D t\D^3x {\cal N}\sqrt{\gamma} {\cal L}_{{\rm m}},
\end{eqnarray}
where ${\cal L}_{{\rm m}}$ is the Lagrangian for matter fields,
\begin{eqnarray}
K_{ij}:=\frac{1}{2{\cal N}}\left(\dot{\gamma}_{ij}-\nabla_i{\cal N}_j
-\nabla_j{\cal N}_i\right)
\end{eqnarray}
is the extrinsic curvature, $R=\gamma^{ij}R_{ij}$ is the trace of the Ricci scalar
(the spatial curvature scalar), and
\begin{eqnarray}
&&{\cal L}_{V 2}:=\alpha_2 R^2+\alpha_3 R_{ij}R^{ij}
+\alpha_4 R^3+\alpha_5 RR_{ij}R^{ij}
\nonumber\\&&\quad
+\alpha_6R_i^{\;j}R_j^{\;k}R_k^{\;i}
+\alpha_7 R\nabla_i\nabla^iR+\alpha_8 \nabla_iR_{jk}\nabla^iR^{jk}.
\end{eqnarray}
The kinetic term coincides with that of general relativity when $\lambda= 1$,
but we do not specify the value of $\lambda$ throughout the paper.
Note however that it has been argued that an additional longitudinal degree of freedom of gravitons
suffers from ghost-like instabilities for $1/3<\lambda<1$~\cite{Horava, Bogdanos:2009uj}.
One may include a cosmological constant in the above action, but we do
not write it explicitly since it can also be thought of as a part of the
matter Lagrangian.
One can also include a parity-violating
term associated with the Cotton tensor,
as in original Ho\v{r}ava gravity~\cite{Horava}.

Variation with respect to ${\cal N}$ yields the Hamiltonian constraint.
In the projectable version of Ho\v{r}ava gravity, the Hamiltonian
constraint is not satisfied locally at each spatial point, but rather a
global equation integrated over the whole space because ${\cal N}$ is a
function of $t$ only.  The global Hamiltonian constraint reads
\begin{eqnarray}
\int\D^3x\sqrt{\gamma}\left[
K_{ij}K^{ij}-\lambda K^2-R-{\cal L}_{V 2}+16\pi G\,E
\right]=0,
\label{gHc}
\end{eqnarray}
where
\begin{eqnarray}
E:=-{\cal L}_{{\rm m}}-{\cal N}\frac{\delta {\cal L}_{{\rm m}}}{\delta{\cal N}}.
\label{Edensity}
\end{eqnarray}
Variation with respect to ${\cal N}_i$ leads to the momentum constraint,
\begin{eqnarray}
\nabla_jP^{ij} = 8\pi G \, J^i,
\end{eqnarray}
where
\begin{eqnarray}
P^{ij}:=K^{ij}-\lambda K\gamma^{ij},
\quad
J^i  = -{\cal N}\frac{\delta{\cal L}_{{\rm m}}}{\delta{\cal N}_i}.
\end{eqnarray}
Finally, variation with
respect to $\gamma_{ij}$ gives the evolution equations,
\begin{eqnarray}
&&2 \left(
K_{ik}K_j^{\;k}-\lambda KK_{ij}
\right)-\frac{1}{2} \left(K_{kl}K^{kl}-\lambda K^2\right)\gamma_{ij}
\nonumber\\&&\quad
+\frac{1}{\cN \sqrt{\gamma}}\gamma_{ik}\gamma_{jl} \partial_t
\left(\sqrt{\gamma}P^{kl}\right)
-\frac{1}{\cN}\nabla^k\left(P_{ij}\cN_k\right)
\nonumber\\&&\qquad
+\frac{1}{\cN}\nabla^k \left(P_{ik}\cN_j\right)
+\frac{1}{\cN} \nabla^k\left(P_{jk}\cN_i\right)
+R_{ij}-\frac{1}{2}R\gamma_{ij}
\nonumber\\&&\qquad\quad
+F_{ij}= 8\pi G\;T_{ij},
\end{eqnarray}
where $F_{ij}:=\delta{\cal L}_{V
2}/\delta\gamma^{ij}-(1/2)\gamma_{ij}{\cal L}_{V 2}$ and
\begin{eqnarray}
T_{ij}:= {\cal L}_m\gamma_{ij}-2\frac{\delta{\cal L}_m}{\delta \gamma^{ij}}.
\end{eqnarray}

The matter action is invariant under the infinitesimal
transformation~(\ref{trans}), which results in the energy-momentum
conservation equations:
\begin{eqnarray}
\int\D^3x\left[
\frac{\sqrt{\gamma}}{2}\dot{\gamma}_{ij}T^{ij}+
\partial_t\left(\sqrt{\gamma} E\right)
+\frac{{\cal N}_i}{{\cal N}}\partial_t\left(
\sqrt{\gamma}J^i\right)
\right]=0,\label{en-cv}
\\
\nabla^jT_{ij}-\frac{1}{\cN\sqrt{\gamma}}\partial_t\left(\sqrt{\gamma} J_i\right)
-\frac{\cN_i}{\cN}\nabla_jJ^j
\qquad\qquad\nonumber\\
-\frac{J^j}{\cN}\left(\nabla_j\cN_i-\nabla_i\cN_j\right)=0.\qquad
\end{eqnarray}
The energy conservation law~(\ref{en-cv}) is of the form of the
integration over the whole space, as is the case for the Hamiltonian
constraint.

\section{Background evolution}

The background evolution of HL cosmology can be derived by setting
${\cal N}=1, \;{\cal N}_i=0$, and $\gamma_{ij} = a^2(t)\delta_{ij}$.
The evolution equation at zeroth order reads
\begin{eqnarray}
\frac{1-3\lambda}{2}\left(3H^2+2\dot H\right) = 8\pi G p,
\label{bg-ij}
\end{eqnarray}
where $H:=\dot a/a$, and $T_{ij} = p\gamma_{ij}$ has been assumed.

Let us {\em define} ${\cal E}(t)$ by
\begin{eqnarray}
8\pi G\left[{\cal E}(t)+\rho\right] =-\frac{3}{2}(1-3\lambda) H^2,
\end{eqnarray}
where $\rho$ is the background value of the matter energy density $E$.
In the case of $\lambda=1$,
the meaning of ${\cal E}$ becomes more transparent by noticing that
$8\pi G{\cal E} = 8\pi G T_0^{\;0}-G_0^{\;0}$, where $G_0^{\;0}$ is the
$(0\,0)$ component of the usual Einstein tensor: ${\cal E}$ arises
because the local Hamiltonian constraint is absent in Ho\v{r}ava
gravity.  This term corresponds to ``dark matter as an integration
constant'' in Refs.~\cite{Mukohyama1, Mukohyama2}.

We emphasize that in this paper the homogeneous background is assumed at
least in our observable patch of the universe because nobody can
tell what happens beyond the present horizon scale.
Under this assumption we can conclude from the global Hamiltonian
constraint that ${\cal E}$ does not necessarily vanish in the local patch. For
example, we may have ${\cal E} >0$ in our patch of the universe, but
${\cal E}$ may be negative in a different patch.  Our assumption is in
contrast to Ref.~\cite{W-M}, in which the standard assumption of a
homogeneous background is made over the whole space, so that the global Hamiltonian
constraint enforces ${\cal E} =0$.  In this
paper, we do not assume the homogeneity over the whole space,
and therefore, Eq.~(\ref{gHc}) does
not constrain the value of ${\cal E}$ in our observable patch of the universe.

In terms of ${\cal E}$, Eq.~(\ref{bg-ij}) can be written in the form of
a conservation equation:
\begin{eqnarray}
\dot{\cal E}+\dot\rho+3H\left({\cal E}+\rho+p\right)=0.
\label{ec-bg}
\end{eqnarray}
This does not guarantee the local conservation of the matter energy
density. If the matter action respects general covariance, we have an
additional conservation equation, $\dot\rho+3H(\rho+p)=0$. 
In the case of scalar field matter~\cite{Kiritsis}, the equation of motion
leads to $\dot\rho+3H(\rho+p)=0$.
Combining the
local conservation of matter energy with Eq.~(\ref{ec-bg}), we obtain
$\dot{\cal E}+3H{\cal E}=0$, implying that ${\cal E}$ indeed shows a
dust-like behavior~\cite{Mukohyama1, Mukohyama2}.

\section{Large scale cosmological perturbations}

Let us study linear perturbations around the cosmological background.
The perturbed metric is given by
\begin{eqnarray}
{\cal N}^2 =  1+2A(t),
\quad
{\cal N}_i= a^2 B_{,i},
\nonumber\\
\gamma_{ij} = a^2\left[(1-2\psi)\delta_{ij}+2D_{,ij}\right].
\end{eqnarray}
Since we are imposing the projectability condition, the perturbation of
the lapse function $A$ does not depend on $\Vec{x}$.  Cosmological
perturbation theory in Ho\v{r}ava gravity without the projectability
condition has been studied in~\cite{GaoBrandenberger}.  Under the scalar
gauge transformation, i.e., the infinitesimal
transformation~(\ref{gtrans}) with $\chi^i =\partial^i\chi(t,\Vec{x})$,
the metric perturbations transform as
\begin{eqnarray}
&&A\to A-\dot \chi^0,\quad
\psi \to \psi +H\chi^0,
\nonumber\\
&&B\to B-\dot\chi,
\quad
D\to D-\chi.
\end{eqnarray}
Since $\chi^0$ depends only on $t$, inhomogeneous $\psi$ cannot be
gauged away, while $A$ can be set to zero by the gauge transformation.
This point is in contrast to general relativity.  It is
convenient to define $\sigma:=\dot D -B$, which is gauge invariant.

The evolution equations take the form
\begin{eqnarray}
&&{\cal G}\,
\delta_i^{\;j}+\left(\partial_i\partial^j-\frac{1}{3}\nabla^2\delta_i^{\;j}\right)
\left[
\frac{\psi }{a^2}+\dot\sigma + 3H\sigma
\right] +\delta F_{i}^{\;j}
\nonumber\\&&\quad
= 8\pi G \,\delta T_{i}^{\;j},  \label{Eq:ij}
\end{eqnarray}
where $\nabla^2:=\delta^{ij}\partial_i\partial_j$,
\begin{eqnarray}
&&{\cal G}:=-(1-3\lambda)\left[\ddot\psi+3H\dot\psi+H\dot A+\left(3H^2+2\dot H\right)A\right]
\nonumber\\&&
+(1-\lambda)\nabla^2\left[\dot\sigma+ 3H\sigma\right]
-\frac{2}{3}\nabla^2
\left[
\frac{\psi }{a^2} +\dot\sigma + 3H\sigma
\right],
\end{eqnarray}
and $\delta F_i^{\;j}$ is to be derived from ${\cal L}_{V 2}$.
Since $\delta F_i^{\;j} ={\cal O}(\nabla^4)$,\footnote{
A straightforward calculation shows that
$R_{ij}=\partial_i\partial_j\psi+\nabla^2\psi\delta_{ij}$
and the variable $D$ does not appear here. Therefore,
$\delta F_i^{\;j}\supset \nabla^4\psi,\nabla^6\psi$.}
it is not important as long as one concerns the large scale evolution
of cosmological perturbations.
(We do not consider the case in which higher spatial derivative terms are
much larger than ${\cal O}(\nabla^2)$ terms, though ${\cal O}(\nabla^6)$ terms
have an interesting effect on the spectrum of perturbations~\cite{Mukohyama-scale-inv}.)
The perturbed energy-momentum tensor may be written
in terms of isotropic and anisotropic pressure perturbations as
\begin{eqnarray}
\delta T_{i}^{\;j}
=\delta p\;\delta_i^{\;j} +
\left(\partial_i\partial^j-\frac{1}{3}\nabla^2\delta_i^{\;j}\right)\Pi.
\end{eqnarray}
The momentum constraint is given by
\begin{eqnarray}
\partial^i\left[ -(1-3\lambda)\dot\psi+(1-\lambda)\nabla^2\sigma\right]
 =8\pi G\,a^2 \delta J^i.
\end{eqnarray}

Analogously to ${\cal E}$ defined in the previous section, let us now
{\em define} $\varepsilon(t,\Vec{x})$ by
\begin{eqnarray}
-8\pi G\left[\varepsilon(t, \Vec{x})+\delta\rho\right] &=&
-3(1-3\lambda)H\left(\dot\psi+HA\right)
\nonumber\\&&\quad
-2\nabla^2
\left(\frac{\psi}{a^2} +H\sigma\right),
\label{def-e}
\end{eqnarray}
where $\delta\rho$ is the perturbation of the matter energy density $E$ given in
Eq.~(\ref{Edensity}). In the case of $\lambda=1$ we have $8\pi
G\varepsilon = 8\pi G\delta T_0^{\;0}-\delta G_0^{\;0}$, from which it
is clear again that $\varepsilon$ is a consequence of the absence of the
local Hamiltonian constraint.  
Thus, $\varepsilon$ may be regarded as a
energy density perturbation of ``dark matter as an integration
constant.''  It is easy to check that $\varepsilon$ transforms as
$\varepsilon\to\varepsilon-\dot{\cal E}\chi^0$ under the gauge
transformation.  In terms of $\varepsilon$, the trace part of the
evolution equations, ${\cal G} =8\pi G\,\delta p$, can be written as
\begin{eqnarray}
&&\dot\varepsilon +\dot{\delta\rho}+3H\left(
\varepsilon +\delta\rho+ \delta p\right)-3\dot\psi\left({\cal E}+\rho +p\right)
\nonumber\\&&
=\frac{1}{8\pi G}\nabla^2\left[2\frac{\dot\psi}{a^2} +2\dot H\sigma
+3H(1-\lambda)\left(\dot\sigma+3H\sigma\right)
\right]
\nonumber\\&&\quad
+{\cal O}\left(\nabla^4\right),
\label{pec}
\end{eqnarray}
which reminds us of the perturbed energy conservation equation.  {\em If
the matter energy density is conserved locally at perturbative order on large scales,}
one has, in addition to Eq.~(\ref{pec}),
\begin{eqnarray}
\dot{\delta\rho}+3H\left(\delta\rho+\delta p\right)
-3\dot\psi(\rho+p) = {\cal O}(\nabla^2).
\label{pecm}
\end{eqnarray}
Note, however, that Eq.~(\ref{pec}) does not necessarily imply the local
conservation of the matter energy density~(\ref{pecm}).

In the following we shall study two different cases ${\cal E}\neq 0$ and
${\cal E}=0$.

\subsection{${\cal E}\neq 0$}

Formally, one can define the following gauge invariant quantities:
\begin{eqnarray}
&&\zeta:= (1-f)\zeta_{{\rm HL}}+f\zeta_{{\rm m}},
\label{def:zeta}
\\
&&\zeta_{{\rm HL}}: =-\psi-H\frac{\varepsilon}{\dot{{\cal E}}},
\quad
\zeta_{\rm m}:=-\psi-H\frac{\delta\rho}{\dot\rho},
\end{eqnarray}
with
\begin{eqnarray}
f(t):=\frac{\dot\rho}{\dot{{\cal E}}+\dot\rho}.\label{deff}
\end{eqnarray}
{\em Only when the matter energy density is conserved locally at zeroth order,} we
may rewrite Eq.~(\ref{deff}) to have
\begin{eqnarray}
f = \frac{\rho+p}{{\cal E}+\rho+p}.
\end{eqnarray}
Using the definition~(\ref{def-e}), $\zeta$ can be written
more explicitly as
\begin{eqnarray}
\zeta =-\psi+\frac{H}{\dot H}\left(\dot\psi +HA\right)
+\frac{2\nabla^2\left(\psi/a^2+H\sigma\right)}{3(1-3\lambda)\dot H},
\label{zeta-metric}
\end{eqnarray}
so that $\zeta$ can be expressed solely in terms of the metric
perturbations.  This is essentially the same as the quantity first
introduced in~\cite{BST}.  Using the variables defined above and
neglecting the ${\cal O}(\nabla^2)$ terms, we can rewrite
Eq.~(\ref{pec}) in a suggestive form as
\begin{eqnarray}
\dot\zeta \simeq -\frac{H}{{\cal E}+\rho+p}\delta p_{{\rm nad}}
+Hc_s^2f(1-f){\cal S}_{\rm HL},
\label{conservation}
\end{eqnarray}
where we have introduced the non-adiabatic pressure perturbation of
matter, $\delta p_{{\rm nad}}:=\delta p-c_s^2\delta\rho$, with
$c_s^2:=\dot p/\dot\rho$,
and the isocurvature fluctuation between
``dark matter as an integration constant'' and ordinary matter,
\begin{eqnarray}
{\cal S}_{\rm HL} := 3 \left(\zeta_{{\rm HL}}-\zeta_{{\rm m}}\right).
\end{eqnarray}
We emphasize that Eq.~(\ref{conservation}) has been derived only by using
the evolution equations.

Equation~(\ref{conservation})
however tells nothing about the large scale evolution of $\zeta$ unless
the evolution of $\zeta_{{\rm m}}$ and $\zeta_{{\rm HL}}$ is specified
(except for the special case $c_s^2f(1-f)\simeq 0$).
If the matter energy is conserved locally, one finds, from
Eqs.~(\ref{ec-bg}) and~(\ref{pecm}), that $\dot f+3Hc_s^2f(1-f) =0$ at
zeroth order and $\dot\zeta_{{\rm m}} \simeq 0$ on large scales at
perturbative order, assuming that $\delta p_{{\rm nad}}=0$.  In this
case it can be shown that $\dot\zeta_{{\rm HL}}\simeq 0$ on large
scales, but $\zeta$ is not conserved in general.  Indeed, it follows
immediately from the definition Eq.~(\ref{def:zeta}) that
$\zeta(t,\Vec{x})=[1-f(t)]\zeta_{\rm
HL}^{(0)}(\Vec{x})+f(t)\zeta^{(0)}_{\rm m}(\Vec{x})$, where
$\zeta^{(0)}_{{\rm HL}} $ and $\zeta^{(0)}_{{\rm m}} $ are the initial
conditions for the corresponding variables.  For dust-like matter with
$\rho\propto a^{-3}$, $f$ is constant since ${\cal E}$ also scales as
$a^{-3}$, and hence $\zeta$ is conserved. This fact was already clear in
Eq.~(\ref{conservation}) with $c_s^2=0$.
Another case in which
$\zeta$ is conserved on large scales is ${\cal S}_{\rm HL} = 0$, that is,
$\zeta_{\rm m}=\zeta_{\rm HL}$.
Whether this ``adiabatic relation''
between ``dark matter'' and usual matter is likely or not depends upon
the specific scenario in the early universe. For example, in
the case discussed in Ref.~\cite{Mukohyama-scale-inv}, ordinary matter
is produced by the decay of the curvaton or the modulus while the
initial condition of ``dark matter'' is determined by quantum
fluctuations of the combination of scalar gravitons
and (real) matter fields.
In this case, there is no
relation between them in general, leading to ${\cal S}_{\rm HL} \ne 0$.

Interesting cases with $f \simeq 0$ and $f \simeq 1$ can be studied
without knowing the evolution of ${\cal S}_{\rm HL}$ and hence
without relying on
the local conservation of the ordinary matter energy density.
If radiation dominates the energy density of the universe at early
times, we have $(1-f) \simeq 0$,
which leads to the conservation of $\zeta$ during that period.
On the other hand, if ``dark matter as an integration
constant'' accounts for a significant portion of
real dark matter and dominates the
energy density of the universe at late times, we have $f \simeq 0$, which again
leads to the conservation of $\zeta$. However, in the intermediate regime,
$f(1-f)={\cal O}(1)$, so that
the curvature perturbation grows provided that ${\cal S}_{\rm HL} \ne
0$.
This property is in accordance with what is found in a conventional multi-fluid
system~\cite{LythWands}.
One should also
notice that in the case where ``dark matter as an integration constant''
constitutes a large portion of real dark matter,
${\cal S}_{\rm HL}$ represents the
isocurvature fluctuation between dark matter and radiation, which is
strongly constrained by the cosmic microwave background (CMB) anisotropy.
Thus, the scenario in which ``dark matter as an integration constant''
is really a dark matter component and there is no natural reason to explain
${\cal S}_{\rm HL} = 0$ gives rise to a large isocurvature fluctuation,
which could be incompatible with the present constraint.

Since $A$ does not depend on $\Vec{x}$, taking the spatial gradient of
Eq.~(\ref{zeta-metric}) yields
\begin{eqnarray}
\partial_t\left(\frac{\partial_i \psi}{H}\right)
\simeq \frac{\dot H}{H^2}\partial_i\zeta.
\label{psi-zeta}
\end{eqnarray}
(One can do essentially the same thing by making the gauge choice $A=0$ instead.)
This equation is useful for reconstructing the curvature perturbation
$\partial_i\psi$ from $\zeta$. 
Note that $\partial_i\psi$ is gauge
invariant because $\psi$ is subject only to the temporal gauge
transformation $t\to t+\chi^0(t)$.  If $\partial_i\zeta$ is constant in
time, one finds $\partial_i\psi = -\partial_i\zeta(\Vec{x}) + H(t)
C_i(\Vec{x})$, where the second term corresponds to the decaying mode.

For an illustrative purpose let us consider conserved matter with the
equation of state $p=w\rho$.  Equation~(\ref{psi-zeta}) is integrated to
give
\begin{eqnarray}
\partial_i\psi(t,\Vec{x}) &=& -\partial_i\zeta_{{\rm HL}}^{(0)}(\Vec{x})
+\partial_i\left[
\zeta_{{\rm HL}}^{(0)}(\Vec{x})-\zeta_{{\rm m}}^{(0)}(\Vec{x})\right]
\nonumber\\&&\times\frac{3(1+w)}{2}
H(t)\int^t\!\!\frac{\D t'}{1+{\cal E}/\rho}.
\end{eqnarray}
For ${\cal E}\ll\rho$, we obtain $\partial_i\psi\simeq
-\partial_i\zeta_{{\rm m}}^{(0)}(\Vec{x})$.

Substituting the curvature perturbation $\partial_i\psi$ into the
traceless part of the evolution equations (\ref{Eq:ij}), we obtain the metric
shear $\partial_i \sigma$. Once these metric perturbations are
determined, the (large scale) CMB anisotropies can be computed by making use of
the perturbed geodesic equations.
The detailed calculation of the CMB temperature anisotropies in HL cosmology
is left for further study.

\subsection{${\cal E}=0$}

In this case it does not make sense to define $\zeta_{{\rm HL}}$, but
still one may define $\zeta$ directly by Eq.~(\ref{zeta-metric}).  The
trace part of the evolution equations implies
\begin{eqnarray}
\dot \zeta\simeq -\frac{H}{\rho+p}\left(\delta p_{{\rm nad}}-c_s^2\varepsilon\right).
\label{nc-E0}
\end{eqnarray}
(Note that $\varepsilon$ is gauge invariant when ${\cal E}=0$.)  Thus,
in general $\zeta$ is not conserved even if $\delta p_{{\rm nad}}=0$.

Let us consider again the simple case where the matter energy is
conserved locally and $\delta p_{{\rm nad}}=0$.  In this case it is easy
to see that $\varepsilon =\varepsilon_0 /a^{3}$.  Integrating
Eq.~(\ref{nc-E0}), we obtain
\begin{eqnarray}
\zeta\simeq \zeta^{(0)}-
\left.\frac{\varepsilon}{3(\rho+p)}\right|_{t=t_0}
+\frac{\varepsilon}{3(\rho+p)},
\end{eqnarray}
where $t_0$ is some initial time.



\section{Conclusions}

In this paper, we have studied the large scale evolution of
the cosmological
curvature perturbation in projectable Ho\v{r}ava gravity,
emphasizing the effect of
``dark matter as an integration constant''~\cite{Mukohyama1, Mukohyama2}
that appears as a consequence of the global Hamiltonian constraint.
Our view is that we cannot tell the cosmological dynamics far outside the present Hubble horizon,
and hence the global Hamiltonian constraint does not provide any information
in our observable patch of the universe.
This assumption makes the impact of ``dark matter as an integration
constant''
rather non-trivial.
The curvature perturbation $\zeta$ has been defined in a gauge invariant manner
with this ``dark matter'' component taken into account.
We then clarified the conditions under which $\zeta$ is conserved on large scales
by invoking the evolution equations.
In particular, we pointed out that
$\zeta$ is sourced by the relative entropy perturbation
${\cal S}_{\rm HL}$ between ``dark matter as an
integration constant'' and ordinary matter.
This source term is effective during the period when $c_s^2 f(1-f)$ is not negligible.
In that period, we need to know the evolution of ${\cal S}_{\rm HL}$
in order to know the evolution of $\zeta$.
This is made possible by assuming the local conservation of the energy density of ordinary matter.

If the ``dark matter'' component constitutes a large portion of
real dark matter, ${\cal S}_{\rm HL}$ corresponds to
the isocurvature fluctuation between radiation and dark matter,
which is strongly constrained by the cosmic microwave background anisotropy.
In this case, one therefore needs a natural reason to explain
${\cal S}_{\rm HL}\simeq 0$,
which is a challenge in HL cosmology.

In the present paper, we have focused on the superhorizon evolution
of the curvature perturbation for a given initial condition.
In order to impose an appropriate initial condition,
we need to specify the scenario of the early stage of the universe and
then to quantize the cosmological perturbations, which would allow us to give
observational prediction for curvature and isocurvature perturbations.
Although the procedure is familiar and established in conventional cosmology,
it is non-trivial in HL cosmology.
It would be interesting to study quantization of
the coupled system of a scalar field and metric perturbations
and solve its evolution from subhorizon (or WKB) to superhorizon regimes in HL cosmology.
Moreover, in HL cosmology
we have a novel mechanism to generate a scale invariant spectrum of quantum fluctuations~\cite{Mukohyama-scale-inv},
which relies on the modified dispersion relation brought by ${\cal O}(\nabla^6)$ terms
and does not require inflation.
A detailed analysis of this mechanism taking into account the effect of the metric perturbations
is yet to be done.
These issues are left for a future study.

\acknowledgments
T.K. and Y.U. are supported by the JSPS under Contact Nos.~19-4199 and
19-720. M.Y. is supported by JSPS Grant-in-Aid for Scientific research
No.\,21740187. 

\end{document}